\documentclass[aps,pre,twocolumn,showpacs]{revtex4-1}
\usepackage{amsfonts}
\usepackage{amsmath}
\usepackage{pgfplots,tikz}
\usepackage{amsthm,graphics,graphicx,subfigure,amssymb,array,color}
\usepackage{epstopdf}
\newcolumntype{C}[1]{>{\centering\arraybackslash}p{#1}}

\usepackage{color}

\renewcommand{\k}{\mathbf{k}} 
\newcommand{\p}{\mathbf{p}} 
\newcommand{\q}{\mathbf{q}} 
\newcommand{\kmax}{k_{\rm max}} 
\newcommand{\kst}{k_{\rm st}} 

\begin{document}

\title{Self-truncation and scaling in Euler-Voigt-$\alpha$ and related fluid models}
\author{Giuseppe Di Molfetta}
\affiliation{LERMA, Observatoire de Paris, PSL Research University, CNRS, Sorbonne Universités, UPMC Univ. Paris 6, UMR 8112, F-75014, Paris France}
\author{Giorgio Krstlulovic}
\affiliation{Laboratoire Lagrange, UMR7293, 
Universit\'e de Nice Sophia-Antipolis, \\
CNRS, Observatoire de la C\^ote dÕAzur, BP 4229,
06304 Nice Cedex 4, France}
\author{Marc Brachet}
\affiliation{Laboratoire de Physique Statistique de l'Ecole Normale 
Sup{\'e}rieure  / PSL Research University, \\
associ{\'e} au CNRS et aux Universit{\'e}s Pierre-et-Marie-Curie Paris 06 et Paris Diderot,\\
24 Rue Lhomond, 75231 Paris, France}
\date{\today}
\begin{abstract}
A generalization of the $3D$ Euler-Voigt-$\alpha$ model is obtained by introducing derivatives of arbitrary order $\beta$ (instead of $2$) in the Helmholtz operator. The $\beta \to \infty$ limit is shown to correspond to Galerkin truncation of the Euler equation. Direct numerical simulations (DNS) of the model are performed with resolutions up to $2048^3$ and Taylor-Green initial data. DNS performed at large $\beta$ demonstrate that this simple classical hydrodynamical model presents a self-truncation behavior, similar to that  previously observed for the Gross-Pitaeveskii equation in Krstulovic and Brachet [Phys. Rev. Lett. 106, 115303 (2011)]. The self-truncation regime of the generalized model is shown to reproduce the behavior of the truncated Euler equation demonstrated in Cichowlas et al. [Phys. Rev. Lett. 95, 264502 (2005)]. The long-time growth of the self-truncation wavenumber $k_{\rm st}$ appears to be self-similar.

Two related $\alpha$-Voigt versions  of the EDQNM model and the Leith model are introduced. These simplified theoretical models are shown to reasonably reproduce intermediate time DNS results. The values of the self-similar exponents of these models are found analytically.

\end{abstract}

\maketitle

\section{Introduction}

Classical Galerkin-truncated systems have been studied since the early 50's in fluid mechanics.
In this context, the (time reversible) Euler equation describing spatially-periodic classical ideal fluids is known to admit, when spectrally truncated at wavenumber $k_{\rm max}$, absolute equilibrium solutions with Gaussian
statistics and equipartition of kinetic energy among all Fourier modes 
\cite{LEE:1952p4100,KRAICHNAN:1955p3039,KRAICHNARH:1973p2909,OrszagHouches}.
Furthermore, the dynamics of convergence toward
equilibrium involves a direct energy cascade toward small-scales and contains (long-lasting) transient that mimic (irreversible) viscous effects
that are produced by the ``gas'' of high-wavenumber
partially-thermalized Fourier modes generating (pseudo) dissipative effects
\cite{Cichowlas:2005p1852,Krstulovic:2009p1876,Krstulovic:2009IJBC}.

In the case of superfluids, the relevant equation is the so-called truncated (or Galerkin-projected) Gross-Pitaevskii equation (TGPE). In the TGPE case, absolute equilibrium can also be obtained by a direct energy cascade, in a way similar to that of the truncated Euler case, with final thermalization accompanied by vortex annihilation. Furthermore, increasing the amount of dispersion produces a slowdown of the energy transfer at small-scales inducing a bottleneck and a partial thermalization that is independent of the truncation wavenumber and takes place below a `self-truncation' wavenumber $k_{\rm st}(t)$ that is observed to slowly increase with time \cite{Krstulovic2011a,Krstulovic2011c,shukla2013}.

The purpose of the present paper is to find and study such self-truncation phenomena in the simpler context of classical hydrodynamics of an ideal fluid. This is obtained by using equation of motion of the Euler type. 
To wit, we study here a simple generalization of the standard $3D$ Euler-Voigt-$\alpha$ 
(non dissipative) model \cite{TitiGWP,TiTiVoigt}. 

Note that various \emph{dissipative} Navier-Stokes-Voigt-$\alpha$ regularizations have been proposed in the last decade as efficient subgrid-scale models in order to address the classical turbulence closure problem, both in hydrodynamics \cite{CaoTiti2009} and in magnetohydrodynamics \cite{Mininni2005}. Results on the regularity properties of this type of dissipative models are given in \cite{KalantarovLevantTiti2009} and the statistical solutions are shown to converge to those of Navier-Stokes (when $\alpha \to 0$) in \cite{RamosTiti2010,TitiLevant2010}. For large values of $\alpha$, dissipative Voigt models are known to inhibit and reduce the transfer of energy to the small scales.

Compared to the Euler equations, the conservative $3D$ Euler-Voigt-$\alpha$ model also penalizes the formation of small scales.
We show that this penalization is enough to produce a self truncation regime.
Our main findings are that the self-truncation regime of this generalized model reproduces the behavior of the truncated Euler equation \cite{Cichowlas:2005p1852}. The long-time behavior of the energy spectrum appears to be self-similar.

To understand this self-similarity we further introduce two different models that are $\alpha$-Voigt versions of the Eddy-Damped Quasi-Normal Markovian (EDQNM) model \cite{Orszag1970,OrszagHouches} and the Leith model \cite{Leith1967}, respectively. Both models are shown to present behaviors that are similar to that of the Euler-Voigt-$\alpha$  model. The relative simplicity of these models allows us to determine the analytical values of the self-similar exponents.

The paper is organized as follows. Section \ref{sec:EVmodel} is devoted to our generalized model. Basic definitions of the Euler-Voigt-$\alpha$ model are given in section \ref{sec:EVdef}. Numerical methods and performed computations are detailed in section \ref{sec:nume}. Our results on the self truncation regime are described in section \ref{sec:st}. The long-time behavior is studied in section \ref{sec:lt}. Related theoretical models are presented in section \ref{sec:tm}. Section \ref{sec:EDQNM} is devoted to the $\alpha$V-EDQNM model and section \ref{sec:Lmodel} to the $\alpha$V-Leith model. Finally, our main results are summarized in section \ref{sec:conclusion} where we give our conclusions.

\section{Euler-Voigt-$\alpha$ model} \label{sec:EVmodel}

\subsection{Definition of the model}\label{sec:EVdef}
The standard $3 D$ Euler-Voigt-$\alpha$ 
model \cite{TitiGWP,TiTiVoigt} is a
partial differential equation 
for the $3D$ velocity field $\textbf{u}(x,y,z,t)$
that explicitly reads
\begin{eqnarray}
(1 -\alpha^2 \nabla^2) \frac{\partial \bf{u}}{\partial t}& =&-({\bf u} \cdot \nabla) {\bf u} - \nabla p \nonumber \\
\nabla \cdot \bf{u} &=& 0.\label{eq:Euler-Voigt}
\end{eqnarray}
The operator $- \alpha ^2 \nabla^2\frac{\partial}{\partial t}$ (as we will see later) suppresses the formation of scales smaller than $\alpha$. The associated wave number to this scales is denoted $k_\alpha=\alpha^{-1}$. We refer to the operator in Eq.\ref{eq:Euler-Voigt} as the $\alpha$-term.

Let us now define the generalized  $3 D$ Euler-Voigt-$\alpha$ model:
\begin{eqnarray}
(1 + (- \alpha ^2 \nabla^2)^{\frac {\beta} {2}})  \frac{\partial \bf{u}}{\partial t} &=&-({\bf u} \cdot \nabla) {\bf u} - \nabla p \nonumber \\
  \nabla \cdot \bf{u} &=& 0,\label{eq:Euler-VoigtG}
\end{eqnarray}
where the power $\beta$ is an even integer. We refer to $\beta$ as the penalization exponent as its increase enhances the suppression of small scale generation. When $\alpha=0$, the generalized model Eq. \eqref{eq:Euler-VoigtG} reduces to the standard $3 D$ incompressible Euler equations
\begin{eqnarray}
 \frac{\partial \bf{u}}{\partial t} + \bf{u} \cdot \nabla \bf{u} = - \nabla p, &\hspace{0.5cm}& \nabla \cdot \bf{u} = 0.\label{eq:Euler}
\end{eqnarray}

We consider here spatially-periodic solutions defined in the domain $\Omega=[0, 2 \pi]^3$. The kinetic energy spectrum $E(k,t)$ associated to \eqref{eq:Euler} is defined as the sum over spherical shells
\begin{equation}
\label{eq:spectrum}
E_0(k,t) = \frac{1}{2} {\displaystyle \sum_{\bf{k} \in \mathbb{Z}^3 \atop k-1/2  < |\bf{k}| < k+1/2 }} |\widehat{\bf{u}}({\bf k},t)|^2,
\end{equation}
and the energy
\begin{equation*}
E_0 =\frac{1}{2 (2 \pi)^3} \int_\Omega {\left|{\bf u}({\bf x},t)\right|^2} d^3x = \frac{1}{2} {\displaystyle \sum_{\bf{k} \in \mathbb{Z}^3}} |\widehat{\bf{u}}({\bf k},t)|^2 ,
\end{equation*}
is independent of time when ${\bf u}$ satisfies the 3D Euler equations (\ref{eq:Euler}).
The conserved energy associated to the generalized Euler-Voigt-$\alpha$ model \eqref{eq:Euler-VoigtG} is straightforward to obtain and reads in physical space
\begin{equation*}
E_\alpha =\frac{1}{2 (2 \pi)^3} \int_\Omega {\bf u} \cdot [1 + (- \alpha ^2 \nabla^2)^{\frac {\beta} {2}}] {\bf u} \, d^3x,
\end{equation*}
and in spectral space
\begin{equation}
E_\alpha = \frac{1}{2} {\displaystyle \sum_{\bf{k} \in \mathbb{Z}^3}} [1+(\alpha k)^{\beta}] |\widehat{\bf{u}}({\bf k},t)|^2 .\label{EQ:defE_aphaTot}
\end{equation}
Consequently, the generalized energy spectrum is defined as
\begin{equation}
\label{eq:spectrum_alpha}
E_\alpha(k,t) = \frac{1}{2} {\displaystyle \sum_{\bf{k} \in \mathbb{Z}^3 \atop k-1/2  < |\bf{k}| < k+1/2 }} [1+(\alpha k)^{\beta}]|\widehat{\bf{u}}({\bf k},t)|^2.
\end{equation}
In the following, we refer to $E_\alpha(k,t)$ as the energy spectrum and $E(k,t)$ as the kinetic energy spectrum.
Equations \eqref{eq:Euler-VoigtG} also conserve the generalized helicity 
\begin{equation}
\label{eq:helicity_alpha}
H_\alpha= \frac{1}{2} {\displaystyle \sum_{}} [1+(\alpha k)^{\beta}] \,\widehat{\bf{u}}({\bf k},t)\cdot\widehat{\bf{\omega}}({-\bf k},t).
\end{equation}
In this work we only consider flows with $H_\alpha=0$.

Let us remark that the differential operator multiplying the r.h.s. of our generalized  $3 D$ Euler-Voigt-$\alpha$ model \eqref{eq:Euler-VoigtG} can be written in Fourier space as $1+(\alpha k)^{\beta}=1+(k/k_\alpha)^{\beta}$. The formal limit $\beta \to \infty$ of Eq. \eqref{eq:Euler-VoigtG} thus corresponds to a standard spherical Galerkin truncation ($\hat {\bf u}({\bf{k})}=0$ for $|{\bf k}|>k_{\rm max}$) of the Euler equation \eqref{eq:Euler} at $k_{\rm max}=k_\alpha$. 
Note that a somewhat similar generalization (but involving the use of a high powers of the Laplacian in the \emph{dissipative} term of forced hydrodynamical equations) have been studied in reference \cite{UrielPRL2008}.

It is well known that the truncated Euler equation admits statistically stationary solutions given by the microcanonical distribution determined by the invariants \cite{LEE:1952p4100,KRAICHNARH:1973p2909}. These solutions are the so called absolute equilibrium and lead to equipartition of energy among Fourier modes. The Euler-Voigt-$\alpha$ model considered as a truncated system, also admits absolute equilibrium solutions. When fully thermalized, Fourier modes can be described by the canonical Gibbs distribution ${\bf \hat{u}}(\k)\sim\mathcal{Z}^{-1} \exp{\{-\beta E_\alpha[{\hat{u}}]\}}$, where $\mathcal{Z}$ is the partition function \footnote{Incompressibility must be taken into account when writing the Gibbs distribution, see \cite{LEE:1952p4100}}. As the invariant energy is quadratic, the absolute equilibrium is Gaussian and the Fourier modes are independent. This leads to the spectra
\begin{eqnarray}
E_\alpha(k)=\frac{3E_\alpha k^2}{ \kmax^3}, &\hspace{.25cm}{\rm or}\hspace{.25cm}& E(k)=\frac{3E_\alpha k^2}{ \kmax^3(1+\alpha^\beta k^\beta)}.\label{Eq:EqAbsEalpha}
\end{eqnarray}
The large scale behavior of the kinetic energy spectrum thus depends on the value of the penalization exponent $\beta$ as $E(k)\sim k^{2-\beta}$. Therefore, in thermal equilibrium the small scales of ${\bf u}$ (i.e. $k\gg k_\alpha$) are penalized when $\beta>0$. 

Taking into account that, in Fourier space, the differential operator in \eqref{eq:Euler-VoigtG} can be defined for real values of $\beta \ge 0$, the choice $\beta = 11/3$ yields an absolute equilibrium $E(k) \sim k^{-5/3}$ and thus a fully thermalized field following Kolmogorov scaling.  In the same vein, choosing in two-dimensions $\beta = 2/3$ also yields Kolmogorov scaling, this time  for the equipartition of enstrophy. This can represent an interesting alternative to the fractal decimation method that was used in reference \cite{PhysRevLett.108.074501}.

\subsection{Numerical method} \label{sec:nume}

The generalized  $3 D$ Euler-Voigt-$\alpha$ equations \eqref{eq:Euler-VoigtG} are solved numerically
using standard \citep{Got-Ors} pseudo-spectral methods
with resolution $N$.
Time marching is performed using a second-order
Runge-Kutta scheme and
the solutions are spherically dealiased by suppressing, at each time step, the
modes for which the wave-vector exceeds
two-thirds of the maximum wave-number $N/2$
(thus a $2048^3$ run is truncated at $|{\bf k}|>k_{\max} = 682$,see {\it e.g.} \cite{BustamanteBrachet2012} for details on the numerical method). 

We consider here solutions of Eq. \eqref{eq:Euler-VoigtG} that correspond to the so-called Taylor-Green (TG) \cite{TG1937} ($2 \pi$-periodic) initial data $\textbf{u}(x,y,z,0)={\textbf u}^{\rm{TG}}(x,y,z)$, with
\begin{equation}
{\bf u}^{\rm{TG}}=(\sin(x) \cos(y) \cos(z),-\cos(x) \sin(y)\cos(z), 0). \label{eq:uTG}
\end{equation}
The simulations reported in this paper were performed using a special purpose symmetric parallel code developed from that described in \cite{PhysRevE.78.066401,PLBMR2010,BustamanteBrachet2012,BBKMPR2013}. The code uses the symmetries of the
Taylor-Green initial data to speed-up computations and optimize memory usage.
The workload for a timestep is (roughly) twice that of a general periodic code running at a quarter of the resolution.
Specifically, at a given computational cost, the ratio of
the largest to the smallest scale available to a computation
with enforced Taylor-Green symmetries is enhanced by a
factor of $4$ in linear resolution. This leads to a factor of $32$ savings in total computational
time and memory usage.
The code is based on FFTW and a hybrid MPI-OpenMP scheme derived from that described in \cite{Mininni:GHOST}. At resolution $2048^3$ we used $512$ MPI processes,
each process spawning $8$ OpenMP threads.

When compared with standard Euler equation \eqref{eq:Euler} runs that were performed in reference \cite{BustamanteBrachet2012},
the only computational advantage of the the generalized  $3 D$ Euler-Voigt-$\alpha$ model  \eqref{eq:Euler-VoigtG}
stems from the much weaker Courant-Friedrichs-Lewy (CFL) condition $U k \, \Delta t<C$ on the time step $\Delta t$ which is conditioned by $k=k_\alpha=\alpha^{-1}$ rather than $k=k_{\rm max}$, when $k_\alpha<<k_{\rm max}$. We have performed a number of high-resolution runs that are summarized in Table \ref{tab:runs}. 
 \begin{table}[ht!] 
 \begin{tabular}{| c | c | c | c | c || c | c | c | c | c ||  }  \hline
Run\ & Res. \ & $\beta$   \ & $ k_\alpha$ &$ t_{\rm max}$\  & Run\ & Res. \ & $\beta$   \ & $ k_\alpha$ &$ t_{\rm max}$\\
   \hline
  Euler & 1024  & $-$ & $-$ & 30  &  11 & 2048 & 2 & $100$    & 15  \\
  \hline
  1  & 1024  & 2 & $20$    & 50  & 12  & 2048 & 2 & $200$    & 15   \\
  2  & 1024 & 2 & $40$    & 50  &  13  & 2048 & 4 & $50$    & 15 \\
  3  & 1024 & 2 & $80$    & 50  & 14  & 2048 & 4 & $100$    & 15  \\
  4  & 1024 & 4 & $20$    & 50  & 15 & 2048 & 4 & $200$    & 15   \\
  5  & 1024 & 4 & $40$    & 50  & 16   & 2048 & 6& $50$  & 15  \\
  6  & 1024 & 4 & $80$    & 50  & 17  &  2048 & 6 & $100$ & 15 \\
  7   & 1024 & 6& $20$  & 50  & 18  &  2048 & 6 & $200$ & 15 \\
  8   & 1024 & 6 & $40$ & 50  & 19  & 512  & 2 & $4$    & 2300   \\
  9   & 1024 & 6 & $80$ & 50  & 20  & 512 & 4 & $4$    & 2300   \\
  10  & 2048  & 2 & $50$    & 15  & 21  & 512 & 6 & $4$    & 2300  \\
   \hline  
    \end{tabular}   \caption{
     List of  runs of the generalized  $3 D$ Euler-Voigt-$\alpha$ model \eqref{eq:Euler-VoigtG} with TaylorGreen initial data \eqref{eq:uTG} and maximum integration time $t_{\rm max}$. }
     \label{tab:runs}   \end{table}
The CFL condition $\Delta t \, k_\alpha \sim 0.1$ was enough to insure both stability and energy conservation up to $0.3\%$ in the worst case.

\subsection{Self truncation} \label{sec:st}

\begin{figure}[h!]
\includegraphics[width=0.99\columnwidth]{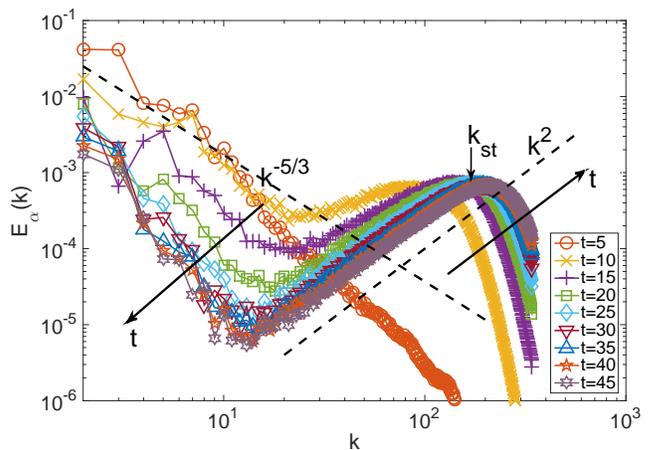}
\caption{(Color online)  Temporal evolution (indicated by arrows) of the energy spectrum $E_\alpha(k)$ for $\beta=4$ and $k_\alpha=80$.  Resolution $1024^3$ ($\kmax=342$). The dashed lines respectively display the Kolmogorov $k^{-5/3}$ and the equipartition $k^2$ scaling. The self-truncation wavenumber is indicated by the small vertical arrow.}\label{fig:EulerSpectraTemporal}
\end{figure}

A first indication on the dynamics of the generalized Euler-Voigt-$\alpha$ model \eqref{eq:Euler-VoigtG} with TaylorGreen initial data \eqref{eq:uTG} is given by the behavior of the Energy spectra for a run at resolution $1024^3$ with $\beta=4$ and $k_\alpha=80$ (run $6$ Table \ref{tab:runs}) that is displayed in Fig.\ref{fig:EulerSpectraTemporal}. 

Different regimes are clearly observed.  First the energy is transferred toward small scales as in the standard Euler equation evolution ($t\leq5$). Then the energy reaches the wavenumber $k_\alpha=80$ and the $\alpha$-term in \eqref{eq:Euler-VoigtG} starts to suppress the energy transfer for $k>k_\alpha$. As a consequence, the energy piles up around this wavenumber, similarly to the truncated Euler case with a cut-off  $\sim k_\alpha$. A compatible Kolmogorov $k^{-5/3}$ scaling is observed at large scales ($t=15$), followed by a partially thermalized zone in $k^2$ extending up to $\kst$. The wavenumber $\kst(t)$ then slowly grows from its initial value $k_\alpha$ until it eventually reaches the simulation cut-off $\kmax$. For $t\to\infty$ the system fully thermalises independently of the parameters (data not shown) and the spectrum is then described by the absolute equilibrium \eqref{Eq:EqAbsEalpha}.
It is conspicuous that for $k>k_{\rm st}$ the energy quickly decays and the partial thermalization regime is thus independent of the simulation cut-off $\kmax$. We thus refer to $k_{\rm st}$ as the \emph{self-truncation} wavenumber. A further discussion and justification for its name will be given later (see below, paragraph following Eq.\ref{eq:kd}).

Figure \ref{fig:EulerSpectra2048} displays the spectra of simulations at resolution $2048^3$ for different values of $\beta$ and $k_\alpha$ taken at time $t=10.2$ \footnote{All along this paper (unless explicitly stated otherwise) colors blue, green and red will be used in plots to denote $\beta=2,4,6$ respectively.}.
\begin{figure}[h!]
\includegraphics[width=0.99\columnwidth]{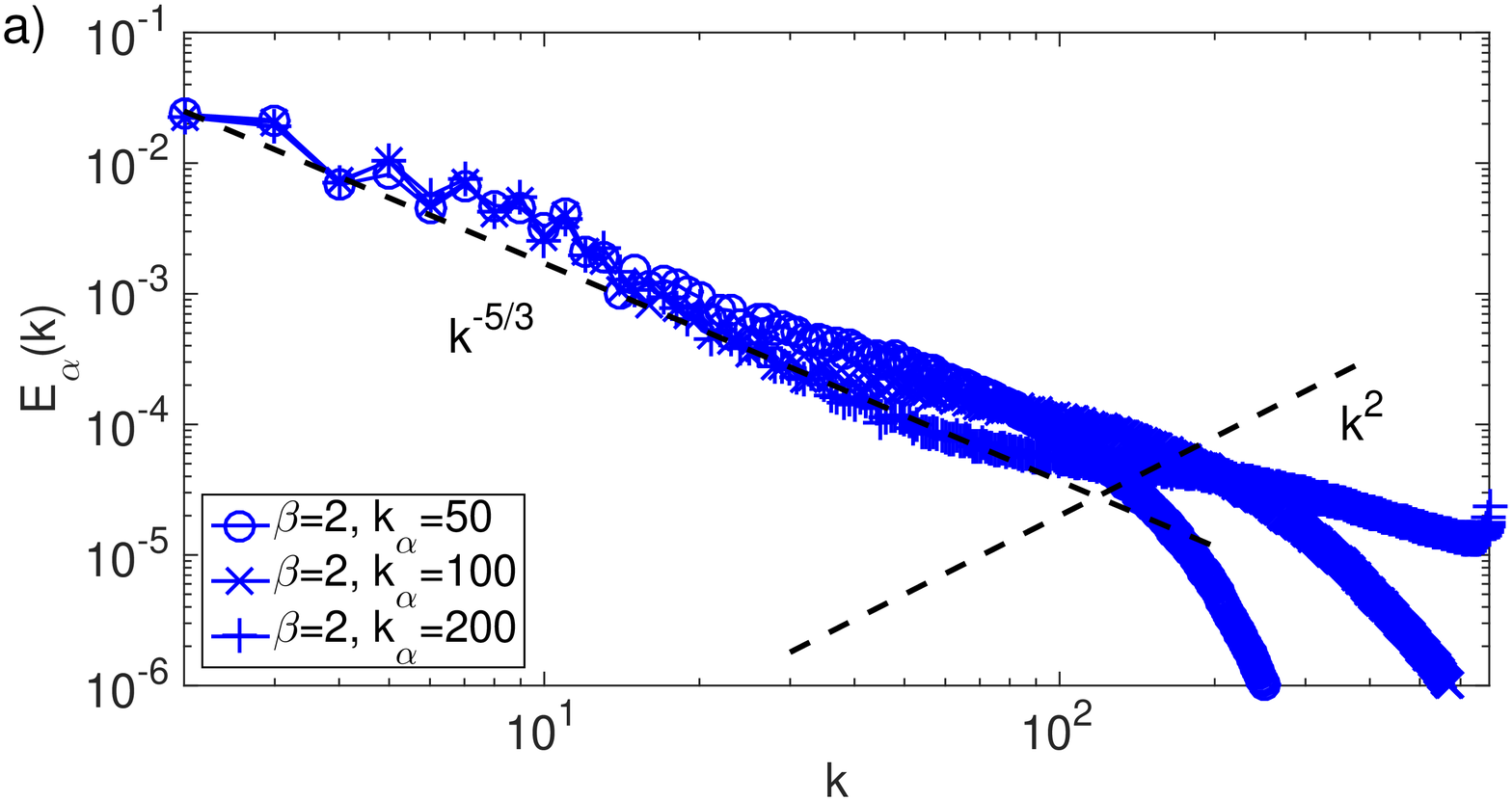}
\includegraphics[width=0.99\columnwidth]{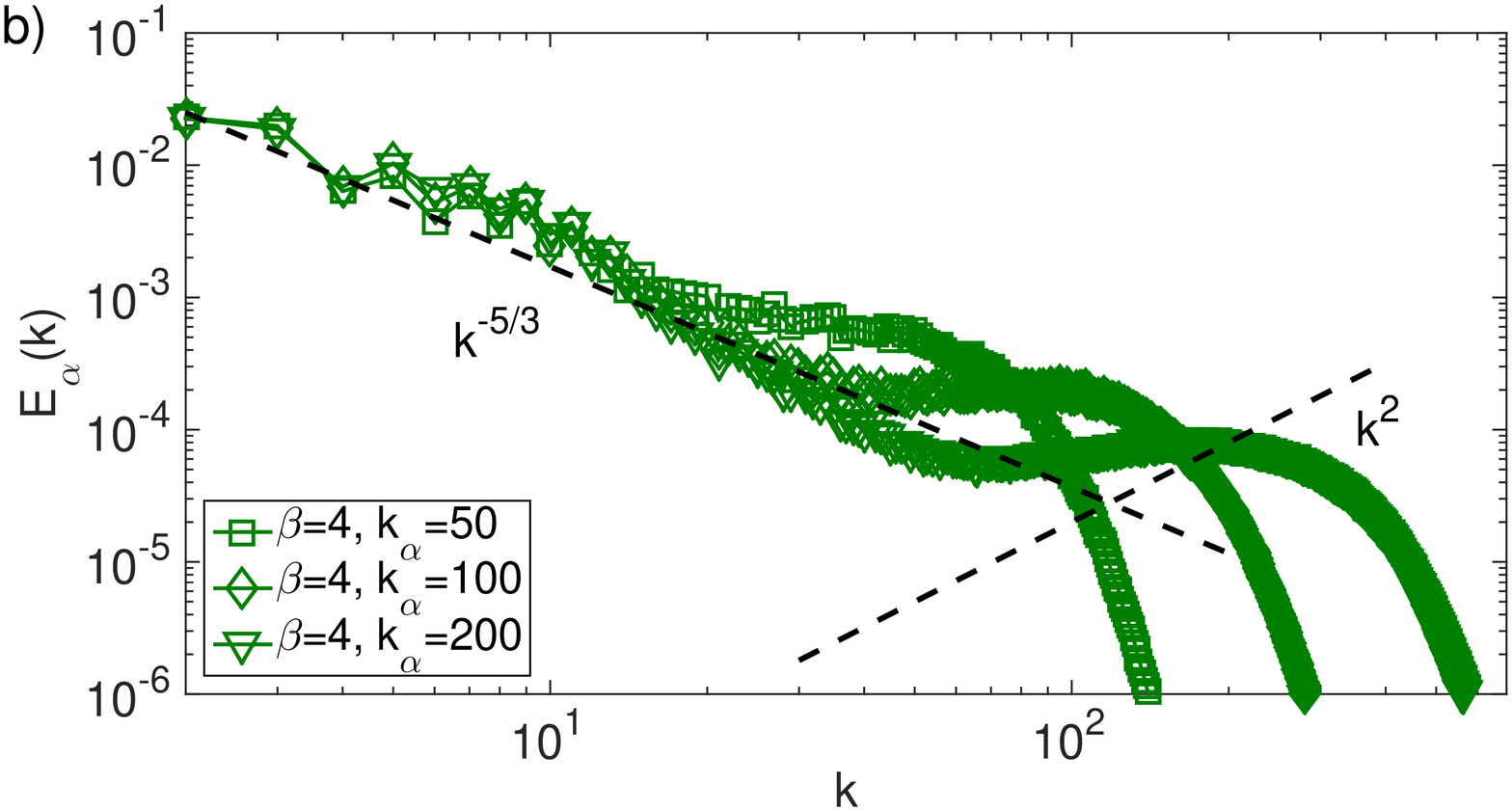}
\includegraphics[width=0.99\columnwidth]{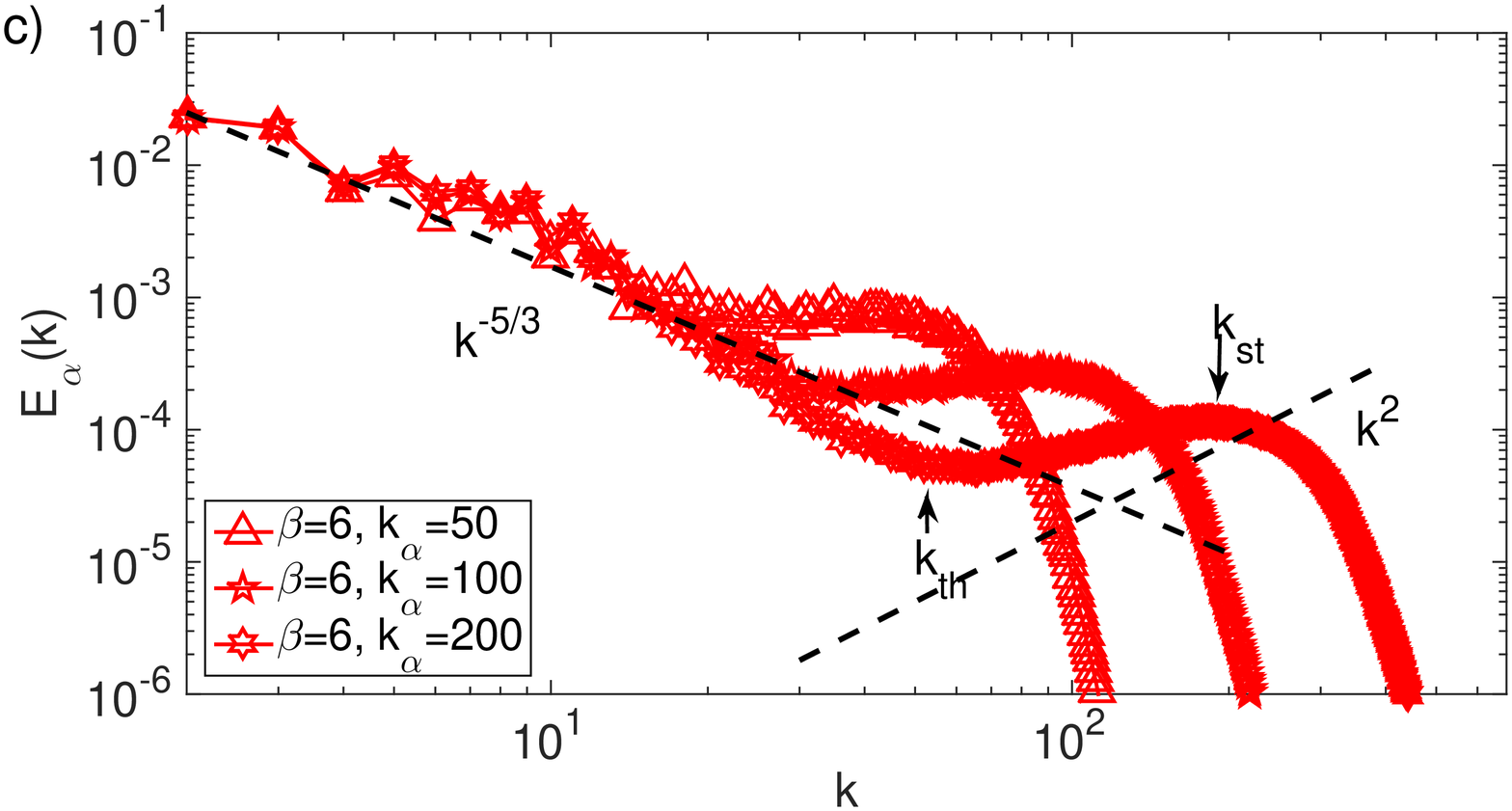}
\caption{(Color online) Energy spectrum $E_\alpha(k)$ versus $k$ for $\beta=2$ and $k_\alpha=50$, $100$ and $200$ at $t=10.2$. (b) Same conditions than in (a) but $\beta=4$. c) Same conditions than in (a) but $\beta=6$. The dashed-black lines corresponds to Kolmogorov scaling $E_\alpha(k)\sim k^{-5/3}$ and energy equipartition $E_\alpha(k)\sim k^{2} $. The thermalization wave number $k_{\rm th}$ and selftruncation $k_{\rm st}$ are indicated by arrows in (c). \label{fig:EulerSpectra2048}}
\end{figure}
The self-truncation is apparent for the smaller $k_\alpha$ at all values of $\beta$ and, for larger $\beta \ge 4$, at all values of $k_\alpha$. The three ranges delimited by the \emph{thermalization} wavenumber $k_{\rm th}$ and the \emph{self-truncation} wavenumber $k_{\rm st}$ are clearly visible (see Fig.\ref{fig:EulerSpectra2048}.c) at this high resolution. Formally, these zones correspond to the Kolomogorov regime ($k\ll k_{\rm th}$), the thermalization range ( $k_{\rm th}\ll k\ll\kst$) and the exponential energy decay ($k\gg \kst$).
It is remarkable that, at this intermediate time ($t=10.2$), the $\beta=6$ case appears to be already behaving somewhat like what is expected of the $\beta=\infty$ limit. Indeed, Fig. \ref{fig:EulerSpectra2048}(c) is reminiscent of the previously studied truncated-Euler case (see Fig.1 of reference \cite{Cichowlas:2005p1852}). Of course, in the  truncated Euler case no third decreasing zone exists as, when $\beta=\infty$, $E_\alpha(k)=0$ for $k>k_{\alpha}$.
In order to make the comparison with the $\beta=\infty$ limit more quantitative, following \cite{Cichowlas:2005p1852}, we have computed the thermalized energy ${E}_{\rm  th}(t) =  \sum_{k_{\rm th}(t) <  k } E(k,t)$ and the effective dissipation $\varepsilon(t)=\frac{{\rm d} {E}_{\rm th}(t)}{{\rm d}t}$. The time evolutions of ${k}_{\rm  th}$, ${E}_{\rm  th}$ and $\varepsilon$ are presented on Fig.\ref{fig:GIU3}.
\begin{figure}[h!]
\includegraphics[width=.99\columnwidth]{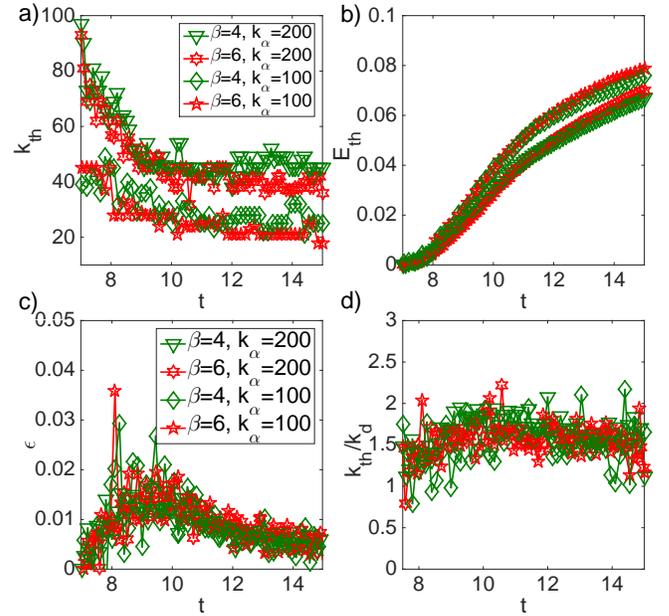}
\caption{(Color online) Temporal evolution of (a) $k_{\rm th}$, (b) $E_{\rm th}$, (d) $\epsilon = \frac{\partial E_{\rm th}}{\partial t}$, (e)  $k_{\rm th}/k_{\rm d}$ with $k_{\rm d}$ estimated based on the self-truncation wavenumber $k_{\rm st}$, see Eq.\eqref{eq:kd}.} 
 \label{fig:GIU3}
\end{figure}
The agreement with the truncated Euler data appears to be  good (compare with Figs 2 and 4 in reference \cite{Cichowlas:2005p1852}). These quantities do not appreciably depend on the value of $\beta$.

In references  \cite{Cichowlas:2005p1852,BertoglioBos}, an effective generalized Navier-Stokes model for the dissipative dynamics of modes $k$ close to $k_{\rm th}(t)$ was suggested for the original Euler case with fixed truncation at $k=k_{\rm max}$. In this case, the effective viscosity of the model was given by $\nu_{\rm eff}=\sqrt{E_{\rm th}}/k_{\rm max}$. If we assume that the generalized large-$\beta$ case behaves similarly to the Euler case truncated at $k_{\rm st}$, we find that the dissipative wavenumber $k_{\rm d}$ should be given by 
\begin{equation} \label{eq:kd}
k_{\rm d} \sim  \varepsilon^{1/4} (\sqrt{E_{\rm th}}/k_{\rm st})^{-3/4} 
\end{equation}
The consistency of this estimation of the effective dissipation with the results displayed in Fig.\ref{fig:GIU3}a-c requires that $k_{\rm d} \sim k_{\rm th}$.
The ratio $k_{\rm th}/k_{\rm d}$ is displayed on Fig.\ref{fig:GIU3}.d. It is indeed of order unity and reasonably constant in time. Thus the large-$\beta$ dynamics of Eq. \eqref{eq:Euler-VoigtG} is seen to emulate the dynamics of the Euler equation, spectrally-truncated at $k_{\rm max}=k_{\rm st}$. For this reason, we call the regime where this behavior takes place the self-truncation regime.

\begin{figure}
\includegraphics[width=0.99\columnwidth]{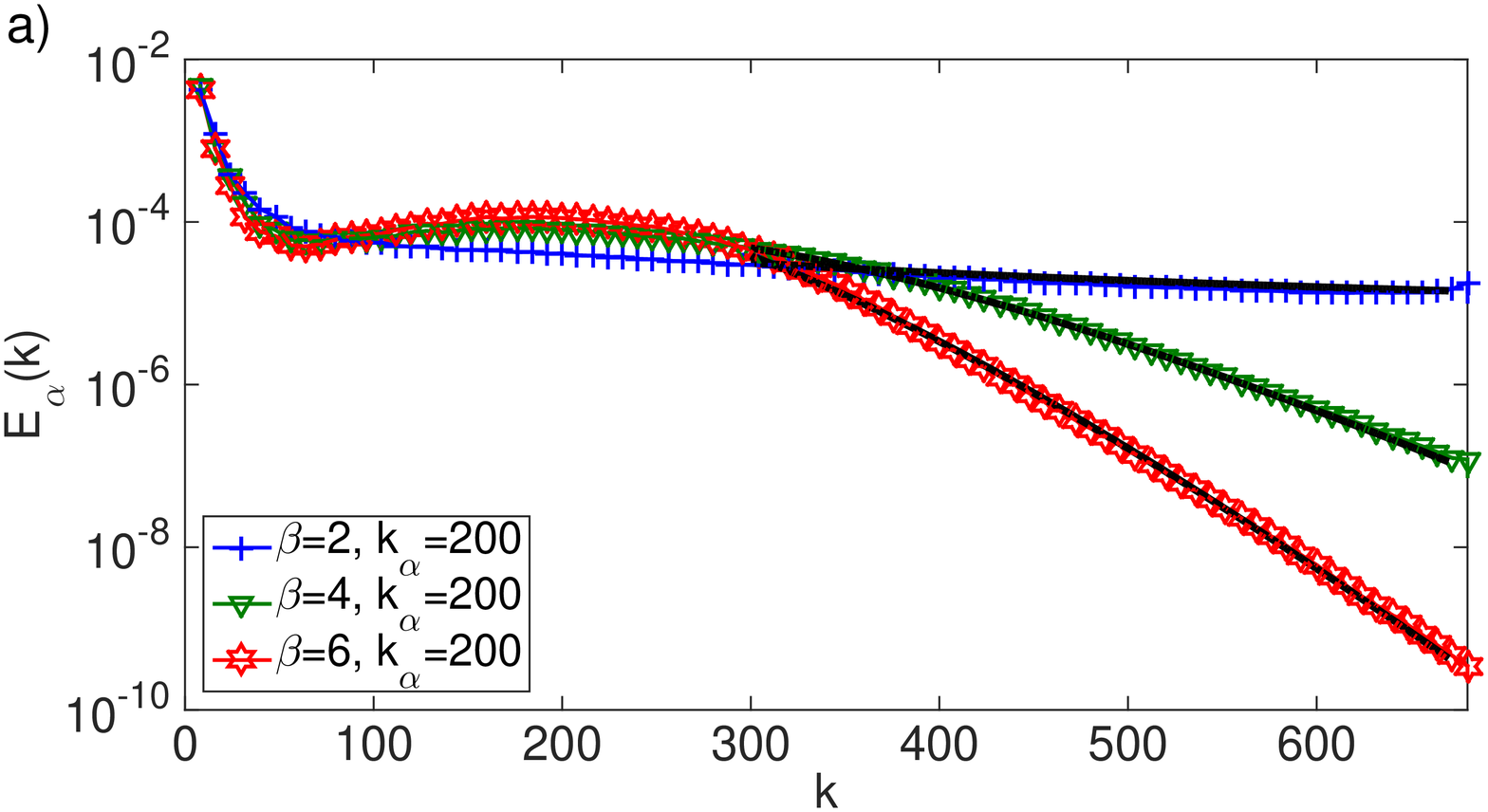}
\includegraphics[width=0.99\columnwidth]{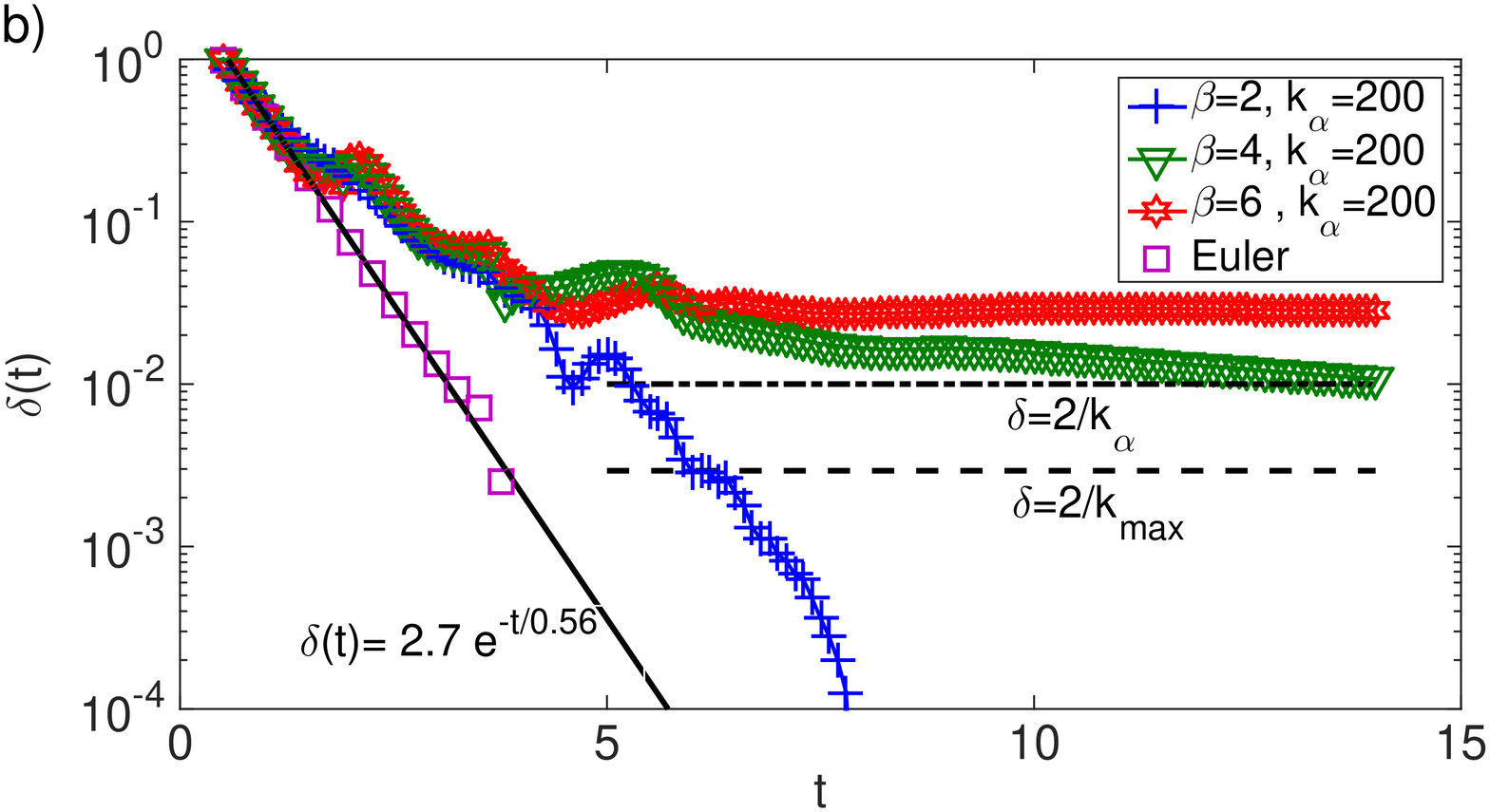}
\caption{(Color online) (a) Exponential decay of  $E_\alpha (k)$ at $t=7.8$. b) Time Evolution of energy spectrum fit parameter $\delta$ of Eq.\eqref{eq_fitTG}: Horizontal lines correspond to $\delta {k_{max}}=2$ (dashed black line) and $\delta {k_{\alpha}}=2$ (dot-dashed black line). Exponential law (dot black line) $\delta$ = $2.7 \exp(-t/0.56)$ from reference \cite{BustamanteBrachet2012}.}
\label{fig:expd}
\end{figure}

In the self-truncation regime the energy spectrum $E_\alpha(k)$ sharply decreases for for $k>k_{\rm st}$ (see Fig.\ref{fig:EulerSpectra2048}). In order to quantify this behavior we have used the so-called analyticity strip (AS) method \cite{SulemSulemFrisch1983}.
Let us recall that the idea is to monitor the `width of the analyticity strip' $\delta (\geq0)$ as a function of time, effectively measuring a `distance to the singularity' \cite{Frisch:notout}. Using spectral methods \cite{Got-Ors}, $\delta(t)$ is obtained directly from the high-wavenumber exponential fall off of the spatial Fourier transform of the solution \cite{frischbook}. The AS method has been used to numerically study putative Euler singularities (see {\it e.g.} reference \cite{BustamanteBrachet2012} for implementation details). 
The common procedure is to perform a least-square fit at each time $t$ on
the logarithm of the energy spectrum $E_\alpha(k,t)$, using the functional
form
\begin{equation}
\label{eq_fitTG}
\ln E_\alpha(k,t) = \ln C(t) - n(t)\, \ln k  - 2 k \,\delta(t).
\end{equation}
Energy spectra are fitted on the intervals $2<k<k^*$ for $t<2$ and on the interval $300 < k < \min(k^*,k_{\max})$ for $t>2$, with $k^*= \inf {\{k \,|\, E(k)< 10^{-32}}\}$ denotes the beginning of round off noise.
The fit presented in Fig.\ref{fig:expd}.a in good agreement with the data. 
The time evolution of the fit parameter $\delta$ is displayed in Fig.\ref{fig:expd}.b, where it is compared with the exponential in time law $\delta(t)=2.7e^{-t/0.56}$ followed by the Euler equation (see reference \cite{BustamanteBrachet2012}). The horizontal lines show the value of the length $2/k_{\max}$ (dashed) and $2/k_\alpha$ (dotted-dashed). It is apparent that the model follows the Euler dynamics as long as $\delta k_\alpha\gg1$. The measure of the fit parameters  is reliable as long as $\delta(t)$ remains larger than a few mesh sizes ($\delta k_{\rm max}\gg1$), a condition required for the smallest scales to be accurately resolved and spectral convergence ensured. Thus the dimensionless quantity $\delta k_{\max}$ is a measure of spectral convergence. Therefore, the self-truncation solutions are solution of the full partial differential equation \eqref{eq:Euler-VoigtG} and not of the spectrally truncated system.

From the mathematical point of view, Eq.\eqref{eq:Euler-VoigtG} can be considered as ordinary differential equation as the right hand side can be shown to be a bounded continuos Lipschitz map between Banach spaces. In this case, existence and regularity can be proved using the same techniques than in (finite dimensional) ordinary differential equations \cite{TitiBardos}.

\subsection{Long-time behavior of $k_{\rm st}$} \label{sec:lt}

\begin{figure}
\centering
\includegraphics[width=0.98\columnwidth]{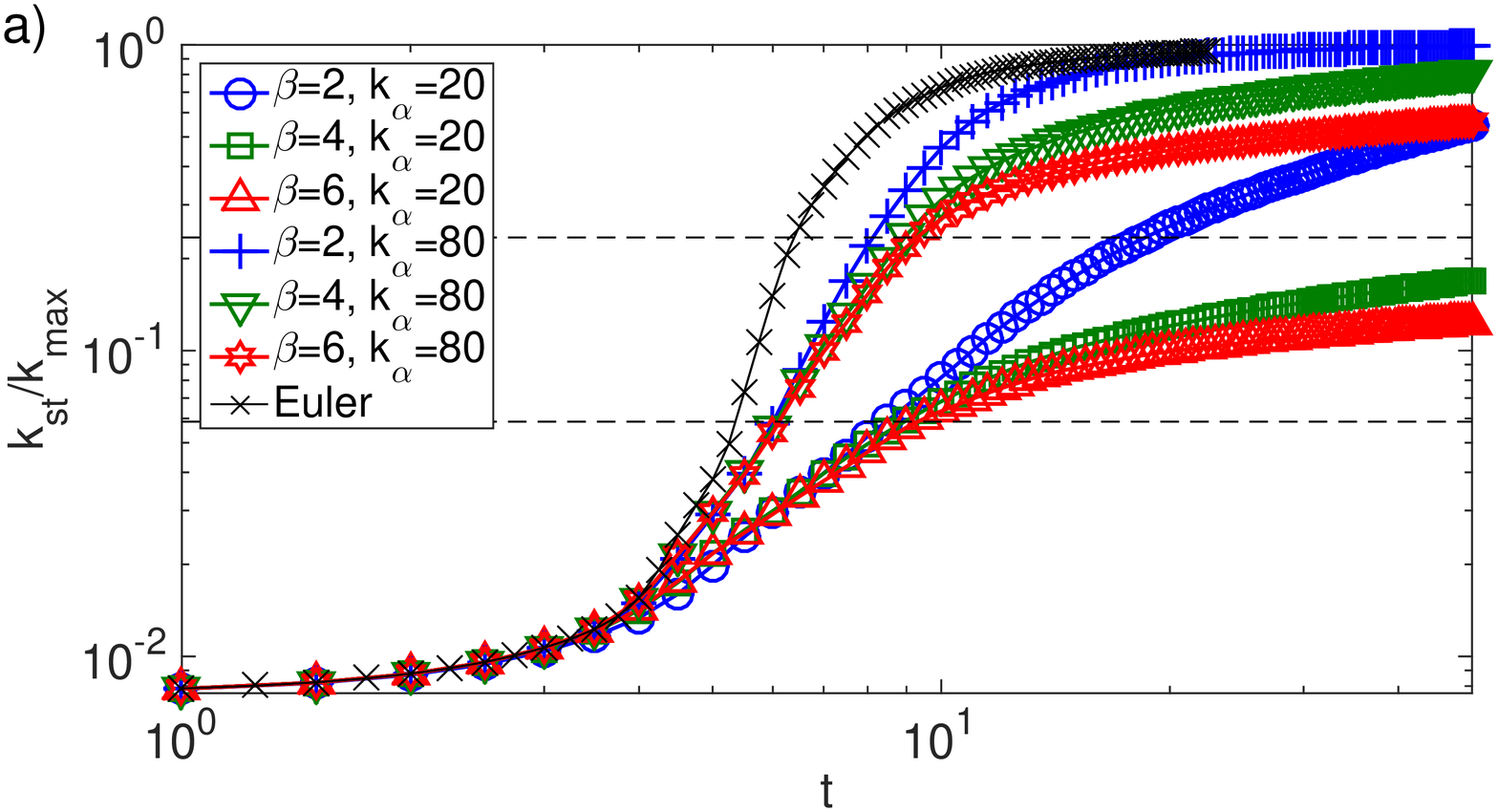}
\includegraphics[width=0.98\columnwidth]{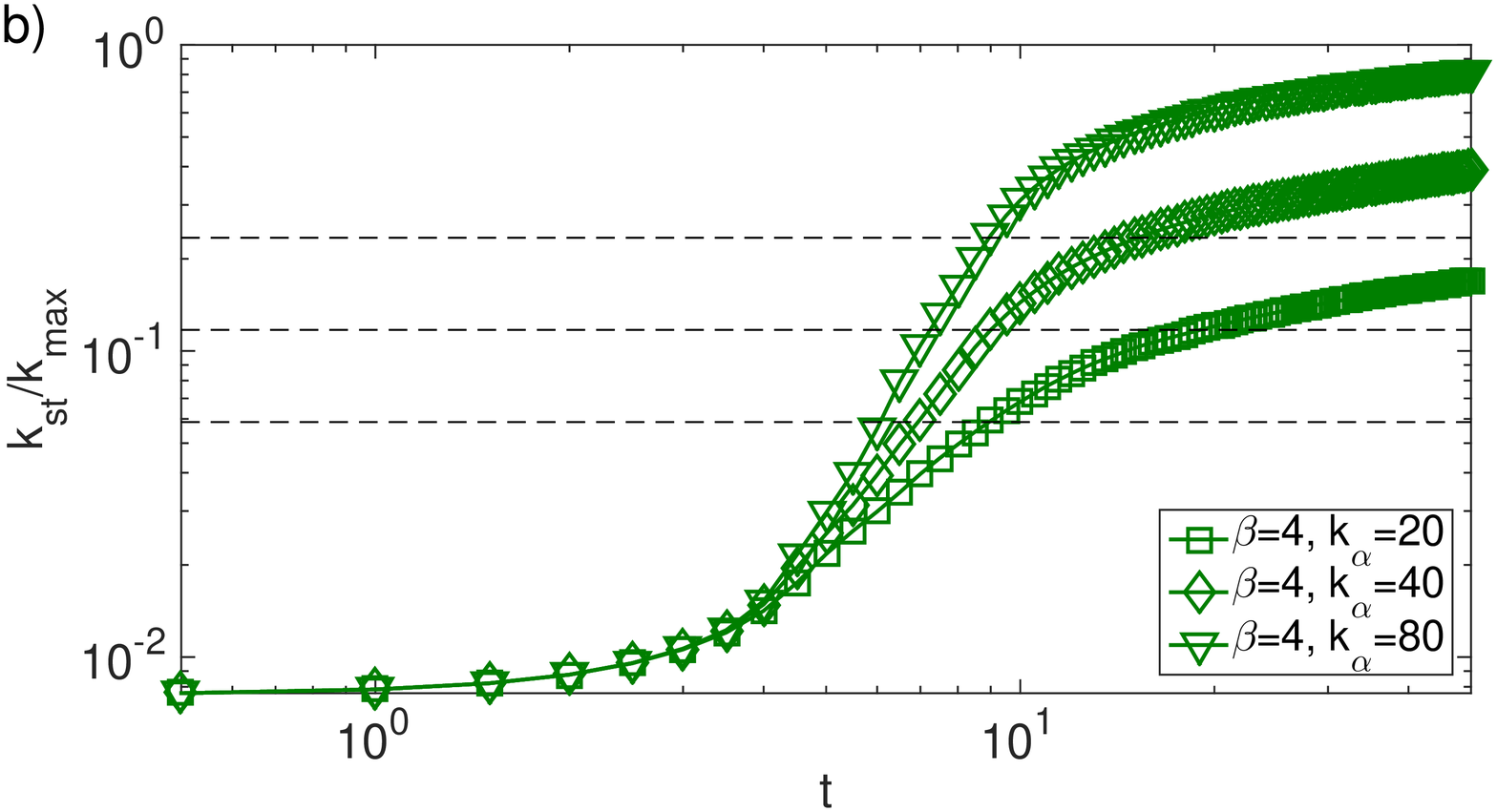}
\caption{(Color online) Time evolution of $k_{\rm st}(t)$ at resolution $1024^3$ (see Table \ref{tab:runs}) . a) Different values of $\beta$ for  $k_\alpha=20$. The crosses show the Euler evolution of $\kst$ ($\alpha=0$). b) $\beta =4$ and different values of $k_\alpha$. The horizontal dashed lines represent the values of $k_\alpha/\kmax$.
}
\label{fig:kst}
\end{figure}

The self-truncation observed in Fig.\ref{fig:EulerSpectra2048} is accompanied by a very slow growth of $\kst$ until it reaches the simulation cut-off $\kmax$. Such a behavior was also observed in the dispersive self-truncation of the truncated Gross-Pitaevskii equation in two \cite{shukla2013} and three \cite{Krstulovic2011a,Krstulovic2011c} dimensions. In the Gross-Pitaevski case, the self-truncation wavenumber $\kst$ was shown to obey a power-law scaling
\begin{equation}
\kst(t)\sim t^\eta,\hspace{1cm} t\gg 1.\label{Eq:DefEta}
\end{equation}
Fig.\ref{fig:kst} displays the temporal evolution of $\kst$ for different values of $\beta$ and $k_\alpha$. 
The wavenumber $k_{st}$ is determined by the weighted average $k_{st} = \sqrt{(5/3)\sum_k  E_\alpha(k,t)k^2/E_\alpha}$. 
With this definition, in the case of the absolute equilibrium \eqref{Eq:EqAbsEalpha} we obtain $\kst=\kmax$. Bear in mind that $\kst$ is defined for all times, even before the self-truncation starts to take place. It is in fact proportional to the inverse of the Taylor microscale \cite{frischbook}.  

Fig. \ref{fig:kst}.a compares $\kst(t)$ for different values of the penalization exponent $\beta$ with $k_\alpha=20$ (lower three curves) and $k_\alpha=80$ (upper three curves). For short times, when $\kst<k_\alpha$, the temporal evolution of $\kst$ is, as expected,  independent of $\beta$. Furthermore, for the very short times, the penalization of small scales introduced by the $\alpha$-term in the Euler-Voigt-$\alpha$ model is negligible and thus $\kst$ is also independent of $k_\alpha$ as the dynamics is given by the (standard) Euler equation (black curve with crosses).

A behavior compatible with a power law is observed at long times. It is apparent that the exponent seems to depend on the value of $\beta$ thought the data do not allow a clear determination of the exponent $\eta$. The power-law behavior is contaminated by the initial dynamic as it is actually expected to have the form $\kst(t)\sim (t-t_0)^\eta$, where $t_0$ is the time when self-truncation starts. Simulations where $t\gg t_0$ and $\kst\ll\kmax$ are difficult to reach with the present choice of parameters and resolution.

In order to explore such power-law behavior a series of runs ($19-21$, see Table \ref{tab:runs}) have been performed in resolution $512^3$ and at small value of $k_\alpha=4$. With this choice of parameter, self-truncation starts at low wavenumbers ($\sim k_\alpha$) and thus no Kolmogorov scaling can be observed. However such a choice allows very long temporal integrations (up to $t=2300$) and to clearly observe the power-law behavior of $\kst$, as apparent in the inset of Fig.\ref{fig:kst}. 
\begin{figure}
\centering
\includegraphics[width=0.99\columnwidth]{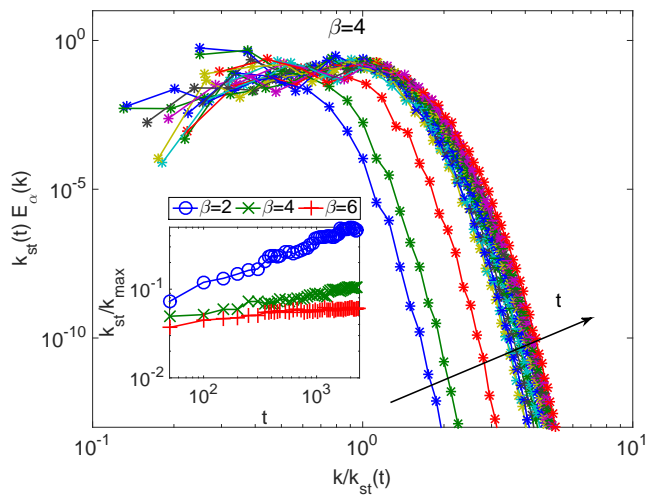}
\caption{(Color online)  a) Temporal evolution of the self-similar function $\Psi(k/k_{\rm st}(t))=E_\alpha(k,t) k_{\rm st}(t)$ (see Eq.\ref{Eq:SelfSimForm}) for $\beta=4$ and $k_\alpha=4$. Data from direct numerical resolution of \eqref{eq:Euler-VoigtG} at resolution $512^3$. The inset shows the temporal evolution of $k_{\rm st}(t)/k_{\rm max}$ for different values of $\beta$. }
\label{fig:kss}
\end{figure}
The value of the exponent clearly depends on the penalization exponent $\beta$. The values measured for $\eta$ are displayed in Table \ref{tab:eta}.
\begin{table}[ht!]
{
 \begin{tabular}{| c || c | c | c |  } \hline
 	 & $\beta=2$ & $\beta=4$ & $\beta=6$  \\ \hline
$\alpha$V-Euler& 	$0.5\pm6\times 10^{-3}$	&  $0.25\pm3\times 10^{-3}$	& $0.07\pm5\times 10^{-3}$ \\ \hline
$\alpha$V-EDQNM  		   	& 	$0.33\pm  5\times 10^{-5}$&  $0.11\pm  9\times 10^{-5}$	&   $0.085\pm  1\times 10^{-4}$\\ \hline
$\alpha$V-Leith		&$0.33\pm10^{-6} $			&   $0.15\pm2\times 10^{-4}$	& $0.09\pm9\times 10^{-6}$  \\ \hline
\end{tabular}   \caption{Values of  the exponent $\eta$ of the self-truncation wavenumber $\kst(t)\sim t^\eta$ (see \eqref{Eq:DefEta}) obtained from direct numerical simulation of the Euler-Voigt-$\alpha$ model \eqref{eq:Euler-VoigtG},  $\alpha$V-EDQNM (\ref{eq_EDQNM}-\ref{eq_eta}) and $\alpha$V-Leith model ($r=2$).\eqref{Eq:LeithModelGen}. }
     \label{tab:eta}  } \end{table}
The power-law observed for $\kst$ strongly suggests to look for self-similar behavior of the energy spectrum $E_\alpha(k,t)$, where the only temporal dependence of the spectrum is given by $\kst(t)$. A self-similar form of the energy spectrum compatible with the conservation law \eqref{EQ:defE_aphaTot} is given by
 \begin{equation}
 E_\alpha(k,t)=\frac{E_0}{k_{\rm st}(t)}\Psi\left(\frac{k}{k_{\rm st}(t)}\right)\label{Eq:SelfSimForm}
 \end{equation}
 Where $E_0$ is a constant with dimension of energy. The function $\Psi(z)$ is expected to behave as $\Psi(z)\sim z^2$ for $z\ll 1$ and exponentially decay for $z\gg 1$. Fig. \ref{fig:kss} displays $\kst(t)E_\alpha(k,t)$ as a function of $k/\kst(t)$ where the tendency to converge towards a self-similar distribution is confirmed for long times. The assumption of self-similarity implicitly supposes that the scale $\alpha$ is very small. Such hypothesis is valid only for long times such that $k_\alpha\ll\kst(t)\ll\kmax$. Discrepancies with the full self-similar form are certainly due to the finite values of the infra-red and ultra-violet cut-off of the simulation. In order to develop further this idea, in the next sections we introduce two theoretical models that allow to obtain both a clear numerical support of self-similarity and an analytic expression for the self-truncation exponent $\eta$. 
 
\section{Theoretical models} \label{sec:tm}

\subsection{Eddy-damped Quasi-Markovian Euler-Voigt-$\alpha$ model}\label{sec:EDQNM}

A popular model of turbulence is the so called Eddy-Damped Quasi-Markovian (EDQNM) closure \cite{Orszag1970,OrszagHouches}. It was derived in the 60-70's and is based on statistical closure of the velocity correlations in Fourier space plus some ad-hoc modeling of the dissipative time scales.  It furnishes an integro-differential equation for the spectrum $E(k,t)$. EDQNM has been proved to be a powerful theoretical and numerical tool in the last 30 years as it allows to achieve a very large scale separation. It was also shown by Bos and Bertoglio \cite{BertoglioBos} that EDQNM reproduces well the dynamics of the truncated Euler equation, including the $k^{-5/3}$ and $k^{2}$ scalings together with the relaxation to equilibrium. It was also used in \cite{Krstulovic2008p2015} to give an analytic prediction of the effective viscosity acting on the large scales of truncated Euler flows.

The extension of EDQNM model to the Euler-Voigt-$\alpha$ case is straightforward. First, following Orszag derivation \cite{OrszagHouches} but using equations \eqref{eq:Euler-VoigtG}, we directly find
\begin{equation}\label{eq_EDQNM}
    \frac{\partial E(k,t)}{\partial t}=\frac{1}{1+\alpha^\beta k^\beta}T_{NL}(k,t)
\end{equation}
where the nonlinear transfer $T_{NL}$ is modeled as
\begin{eqnarray}
\nonumber T_{NL}(k,t)&=&\int\int\limits_\triangle\Theta_{kpq}(xy+z^3)\left[\frac{k^2pE(p,t)E(q,t)}{1+\alpha^\beta k^\beta}\right.\\
&-&\left.\frac{p^3E(q,t)E(k,t)}{1+\alpha^\beta p^\beta}\right]\frac{dp\,dq}{pq} \label{eq_Tnl}.
\end{eqnarray}
In (\ref{eq_Tnl}), $\triangle$ represents a
strip in $(p,q$) space such that the three wavevectors $\k$, $\p$,
$\q$ form a triangle. $x$, $y$, $z$, are the cosine of the angles
opposite to $\k$, $\p$, $\q$. $\Theta_{kpq}$ is a
characteristic time defined as
\begin{equation}\label{eq_Theta}
    \Theta_{kpq}=\frac{1-\exp{(-(\eta_k+\eta_p+\eta_q)t)}}{\eta_k+\eta_p+\eta_q}.
\end{equation}
The standard EDQNM equations are recovered by setting $\alpha=0$. Absolute equilibrium \eqref{Eq:EqAbsEalpha} is a stationary solution of \eqref{eq_EDQNM} that satisfies detailed balance (no flux solution). The eddy damped inverse time $\eta_k$ is defined as
\begin{equation}\label{eq_eta}
    \eta_k=\lambda'\sqrt{\int\limits_0^k \frac{s^2E(s,t)}{1+\alpha^\beta s^\beta}\,ds}.
\end{equation}
In the limit $\alpha\to 0$ the standard eddy damped inverse time is recovered \cite{Pouquet1975evolution}. Note that for $\beta\to\infty$, the standard eddy time is also recovered for $k<k_\alpha=1/\alpha$ whereas $\eta_k=0$ for $k>k_\alpha$, consistently with the dynamics of the truncated Euler equation, as for $k>\kmax$ the dynamics is frozen.
The constant $\lambda$ defines a time scale and we use the standard value $\lambda=0.36$. The truncation is imposed by omitting
all interactions involving waves numbers larger than $k_{\rm{max}}$ in (\ref{eq_Tnl}). 
The main difference between Eqs.(\ref{eq_EDQNM}-\ref{eq_eta}) and the standard EDQNM model comes from the penalization terms that inhibit the energy transfer and modifies the straining. Indeed, Eqs. (\ref{eq_EDQNM}-\ref{eq_Tnl}) can be rewritten as the standard EDQNM equations for the generalized spectrum $E_\alpha$ by redefining the characteristic time \eqref{eq_Theta} as $ \widetilde{ \Theta}_{kpq}=    \Theta_{kpq}/(1+\alpha^\beta k^\beta)(1+\alpha^\beta p^\beta)(1+\alpha^\beta q^\beta)$. The evolution of $E_\alpha$ is thus like the one of $E$ in the standard EDQNM model but with a local modification of the energy transfer term.
We refer to (\ref{eq_EDQNM}-\ref{eq_eta}) as $\alpha$V-EDQNM model. 

\begin{figure}[h!]
\includegraphics[width=0.99\columnwidth]{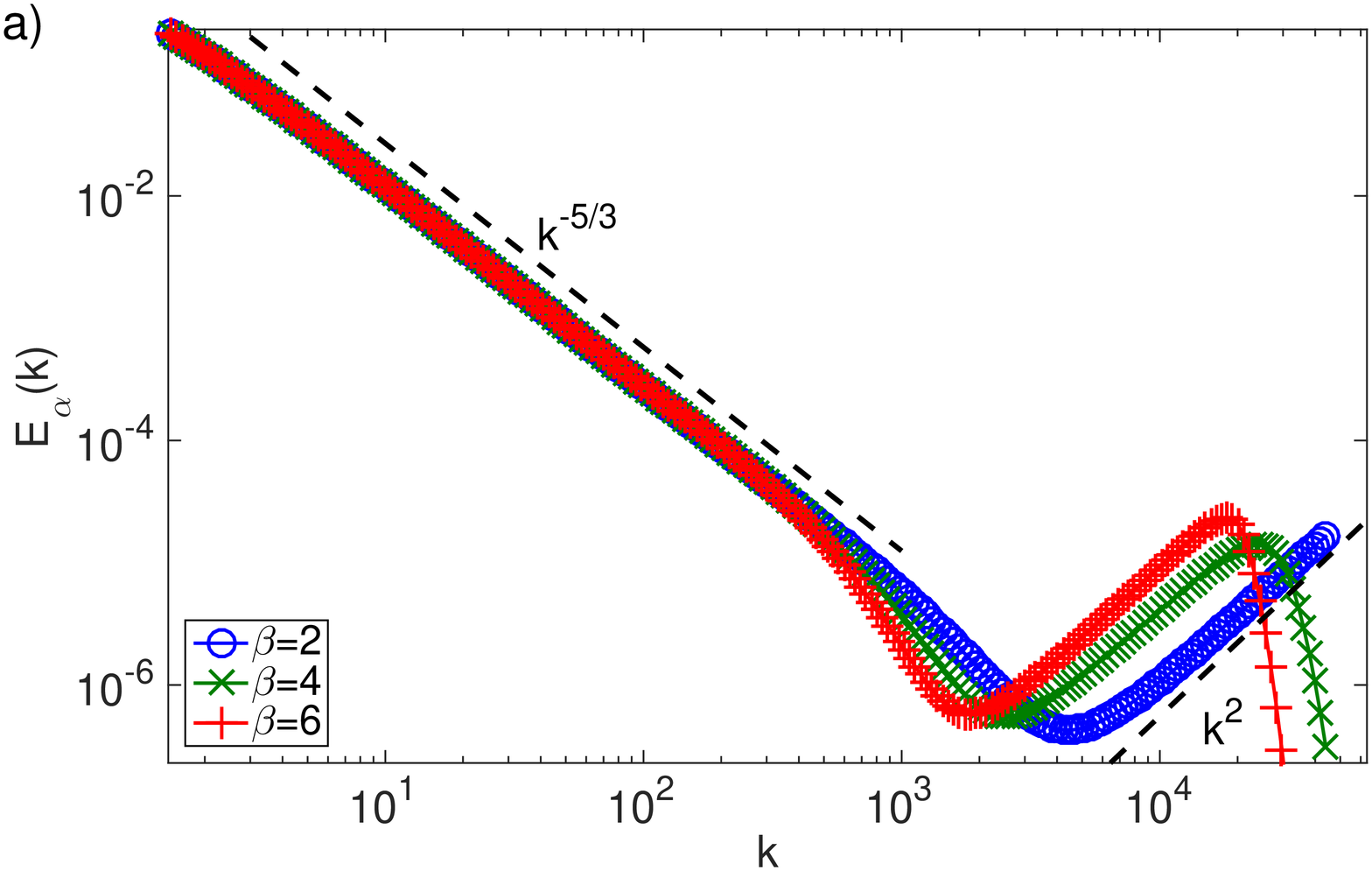}
\includegraphics[width=0.99\columnwidth]{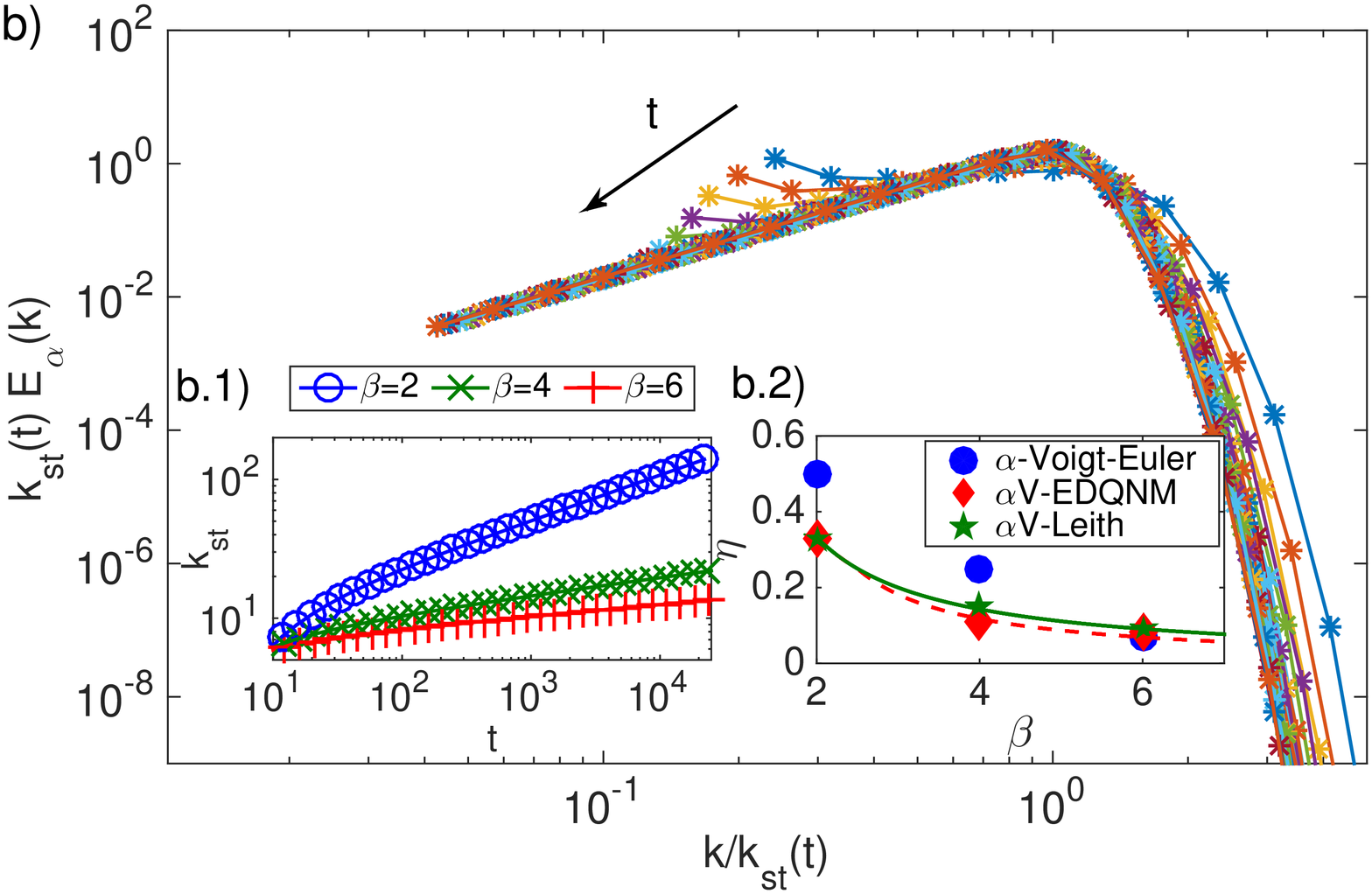}
\caption{(Color online) a) Energy spectra $E_\alpha(k)$ at $t=3.5$ of the $\alpha$V-EDQNM model, for different values of $\beta$, obtained with $k_\alpha=10^4$ and $k_{max}=43918$ (corresponding to resolution $131072$, with $14.2$ points per octave). b) Temporal evolution of the self-similar function $\Psi(k/k_{\rm st}(t))=E_\alpha(k,t) k_{\rm st}(t)$ for $\beta=4$ and  $k_\alpha=5$ and $k_{max}=692$ ($28.4$ points par octave). The inset b.1 shows the temporal evolution of $k_{\rm st}(t)$ for different values of $\beta$. The inset b.2 shows the $\alpha$V-EDQNM (red dashed line) and $\alpha$V-Leith (solid green lines) theoretical predictions \eqref{Eq:Eta_EDQNM} and \eqref{Eq:Eta_Leith} for the self-truncation exponent $\eta$ and numerical data from table \ref{tab:eta}.}  
\label{Fig:EDQNM}
\end{figure}

A number of simulations of the $\alpha$V-EDQNM equation has been performed and give a behavior that is similar to that of the DNS of the full Euler-Voigt-$\alpha$ model, including comparable time scales. Fig. \ref{Fig:EDQNM}.a displays the energy spectrum for $\beta=2,4,6$ and $k_\alpha=10^4$ for a simulation with $\kmax=43918$ (and $14.2$ points per octave).

Fig. \ref{Fig:EDQNM}.a shows that the three zones observed in the Euler-Voigt-$\alpha$ model (see Fig.\ref{fig:EulerSpectra2048}), are also apparent in $\alpha$V-EDQNM model spectra, with scaling laws extending for more than two decades. In the same spirit than in the Euler-Voigt-$\alpha$ DNS runs with a smaller value of $k_\alpha=5$ have been performed, allowing a clear determination of the self truncation exponent $\eta$. The values are presented in Table \ref{tab:eta}. Analogously to the Euler-Voigt-$\alpha$ DNS, we look for a self-similar behavior of the energy spectrum. The collapse is manifest in Fig. \ref{Fig:EDQNM}.b where the self-similar form \eqref{Eq:SelfSimForm} is displayed for $\beta=4$.
The $\alpha$V-EDQNM model allows for directly looking for such self similar behavior and find the exponent $\eta$ by counting powers. Indeed, introducing the self-similar form  and the variable $z=k/\kst(t)$ in \eqref{Eq:SelfSimForm} in the eddy damped inverse time \eqref{eq_eta} we find 
 \begin{equation}
    \eta_z(t)=\lambda E_0^{1/2} k_{\rm st}(t)\sqrt{\int_0^{z}\frac{\Psi\left(z'\right) z'^2 }{(1+\alpha^\beta k_{\rm st}(t)^\beta z'^\beta)^2}\,dz'}.
\end{equation}
The eddy damped inverse time can thus be expressed as $\eta_z=\lambda E_0^{1/2} k_{\rm st}I(z,\alpha \kst)$.  Assuming that $I(z,\alpha\kst)\sim (\alpha\kst)^\gamma$ we obtain for the characteristic time in the self-similar form $\Theta_{z_kz_pz_q}\sim \kst^{-1-\gamma}$. In the same way, the non-linear transfer term \eqref{eq_Tnl} is found to scale with the self-truncation number as
 $T_{NL}\sim k_{\rm st}^{-1-\gamma+3-2-3\beta}$. Equating the left and right hand sides of \eqref{eq_EDQNM} we obtain $\dot{\kst}\kst^{-2}\sim k_{\rm st}^{-\gamma-3\beta}$. Finally,  the scaling $\kst(t)\sim t^\eta$ leads to
\begin{equation}
\eta=\frac{1}{3\beta -1+\gamma}\label{Eq:Eta_EDQNM}
\end{equation}
The exponent $\gamma$ can be computed using the ansatz $\Psi(z)=z^2 \exp{[-z]}$ and it reads $\gamma=-\min{[\beta,5/2]}$. The theoretical prediction confronted with the data are presented in good agreement in the inset b.2 of Fig.\ref{Fig:EDQNM}.

\subsection{$\alpha$-Voigt Leith model.} \label{sec:Lmodel}

Let us now introduce another model that shares the same dynamics properties than Euler-Voigt-$\alpha$. It is a spectral diffusion model that generalizes the so-called Leith model \cite{Leith1967}. The original Leith model it is a phenomenological non-linear (local) spectral diffusion equation for the energy spectrum that admits the stationary solutions corresponding to an absolute equilibrium and Kolmogorov scaling. When forcing and dissipation is added to the model, a steady state containing mixture of constant flux and thermal equilibrium was observed in \cite{WarmCascade}. It also known to posses self-similar solutions \cite{Grebenev_Leith}.

The simplest generalization of the Leith model to take into account the $\alpha$-term that conserves the total energy $E_\alpha(k)$ is given by
\begin{equation}
 \frac{\partial E}{\partial t}=-\frac{1}{(1+\alpha^\beta k^\beta)} \frac{\partial F}{\partial k},
\end{equation}
where $F(k)$ is a spectral (non-linear) flux. Following Leith's original derivation, we assume that the spectral flux is defined in terms of a diffusion coefficient $D(k)$ and a potential $Q(k)$ such that
\begin{equation}
F(k)=-\gamma k^2D\frac{\partial Q}{\partial k}
\end{equation}

Assuming locality in the flux, by dimensional analysis we obtain
\begin{eqnarray}
D&=& k^{9/2-m}(E/k^2)^n (1+\alpha^\beta k^\beta)^p\\
Q&=& k^m (E/k^2)^{3/2-n}(1+\alpha^\beta k^\beta)^q
\end{eqnarray}
The dimensionless coefficient $\gamma$ sets the global time-scale and  $n,m,q,p$ are free parameters to be determined. The first contraint is given by the no-flux solution or the absolute equilibrium \eqref{Eq:EqAbsEalpha}. Imposing that $F(k)=0$ for the absolute equilibrium $E_\alpha(k)\sim k^2$ leads to $m=0$ and $n=3/2-q$. The second contraint is for the Kolmogorov-like solution $E_\alpha\sim k^{-5/3}$. Imposing that for such solutions the flux is given by $F=\epsilon/(1+\alpha^\beta k^\beta)^r$, we obtain $p=3/2-q-r$, being $r$ a free parameter. Such a flux can be easily interpreted in the limit of large $\beta$. Indeed, for $k\ll k_\alpha$ the $\alpha$ term is negligible, and the flux becomes constant. The  Kolmogorov phenomenology is thus recovered. On the other hand for $k\gg k_\alpha$ the flux vanishes as expected. Taking into account the previous constraints, we obtain a family of a diffusive $\alpha$V-Leith models indexed by the parameter $r$:
\begin{equation}
 \frac{\partial E_\alpha}{\partial t}=\frac{2q\gamma }{3}\frac{\partial}{\partial k}\left[\frac{ k^{13/2} }{( 1+\alpha^\beta k^\beta)^r}\frac{\partial}{\partial k}\left[  E_\alpha^{3/2} k^{-3} \right]\right] \label{Eq:LeithModelGen}.
\end{equation}
The standard Leith model is recovered by setting $\beta=0$ and rescaling the time. Note that Eq.\eqref{Eq:LeithModelGen} can be reinterpreted as a standard nonlinear diffusion equation for the generalized energy spectrum $E_\alpha$ by introducing the triple decay time $\tau_3^\alpha(k)$ (see e.g. ref. \cite{Zhou1990}). In terms of $\tau_3^\alpha(k)$ Eq.\eqref{Eq:LeithModelGen} reads
\begin{equation}
 \frac{\partial E_\alpha}{\partial t}=\frac{2q\gamma }{3}\frac{\partial}{\partial k}\left[\tau_3^\alpha(k)k^7\,E_\alpha\frac{\partial}{\partial k}\left[ k^{-2}  E_\alpha \right]\right] \label{Eq:LeithModelGenTripleDecay},
\end{equation}
where the triple decay time is given by $\tau_3^\alpha(k)=\tau_{\rm NL}^\alpha(k)/( 1+\alpha^\beta k^\beta)^r$ with the eddy turnover time defined as $\tau_{\rm NL}^\alpha(k)=[k^{3/2}E_\alpha^{1/2}]^{-1}$ \footnote{In the literature, the spectral transfer time (denoted here by $\tau_{\rm S}^\alpha$) usually appears; it is related to the triple decay time by the relationship $\tau_3^\alpha\,\tau_{\rm S}^\alpha={\tau_{\rm NL}^\alpha}^2$}. Different choices of $\tau_3^\alpha$ could be used to model for instance in magnetohydrodynamics flows the presence of Alfv\'en waves \cite{Zhou1990}.

\begin{figure}[h!]
\includegraphics[width=0.95\columnwidth]{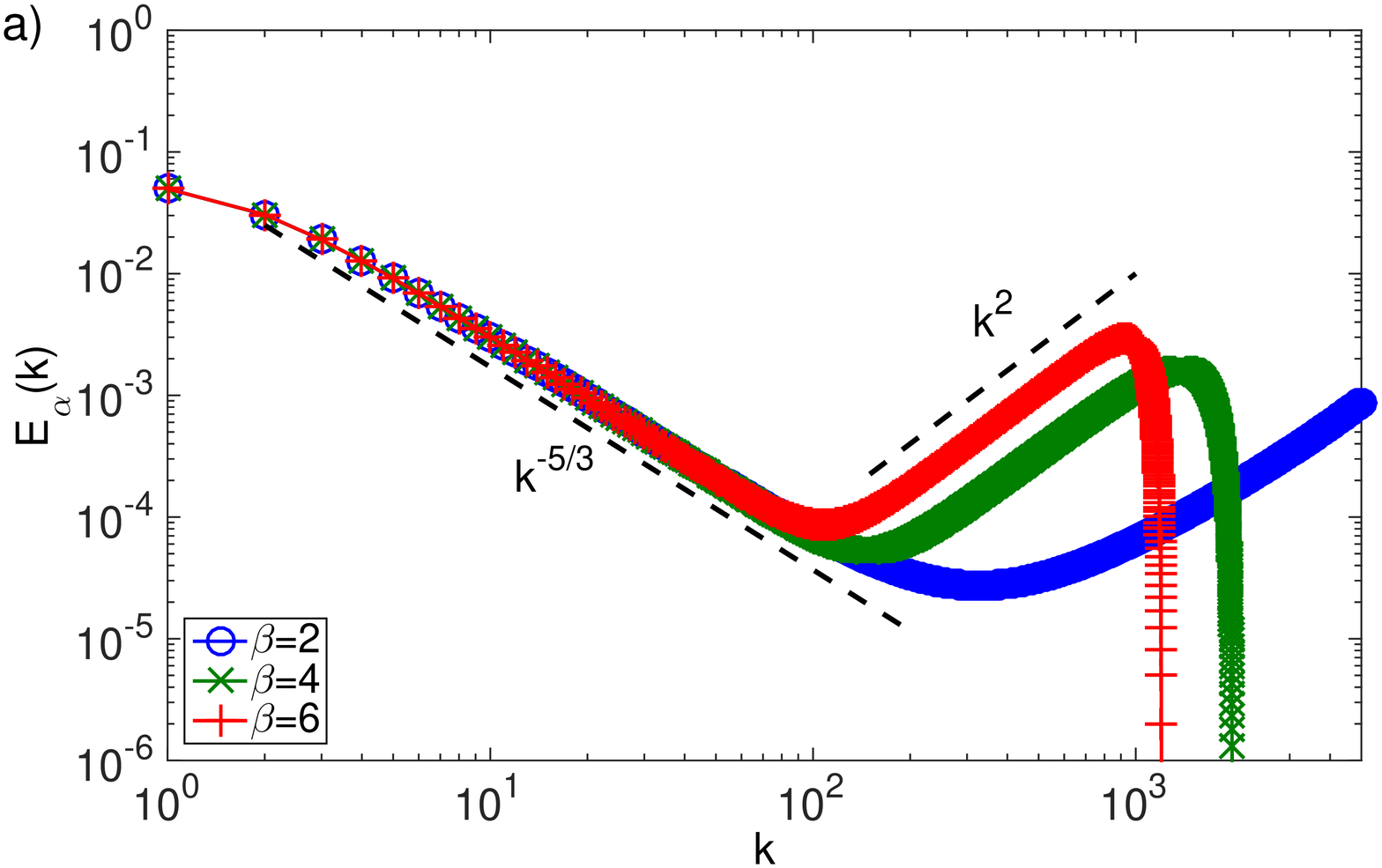}
\includegraphics[width=0.95\columnwidth]{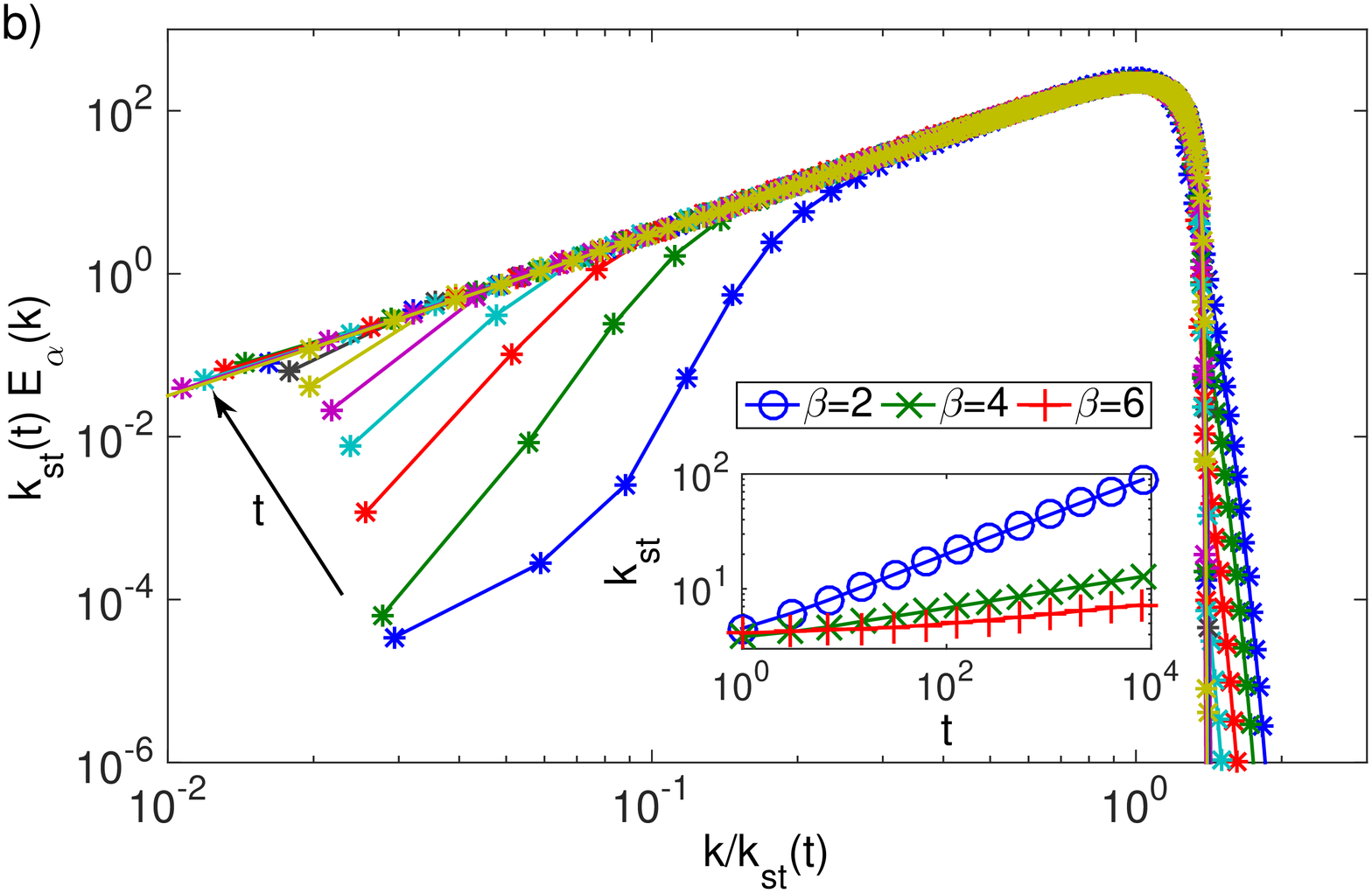}
\caption{(Color online) a) Energy spectra $E_\alpha(k)$ of the $\alpha$V-Leith model \eqref{Eq:LeithModelGen} for different values of $\beta$ obtained with $k_\alpha=400$ and $k_{max}=4000$. b) Temporal evolution of the self similar function $\Psi(k/k_{\rm st}(t))=E_\alpha(k,t) k_{\rm st}(t)$ for $\beta=4$ and $k_\alpha=2$. The inset shows the temporal evolution of $k_{\rm st}(t)$ for different values of $\beta$.\label{FIG:Leith}}
\end{figure}

The solutions of the $\alpha$V-Leith model \eqref{Eq:LeithModelGen} indeed reproduce the ones of  the Euler-Voigt-$\alpha$ equations. Figure \ref{FIG:Leith}.a displays the energy spectrum for different values of $\beta$ and $r=2$. The Kolmogorov $k^{-5/3}$, the equilibrium $k^2$ regimes and the fast decay for large $k$ is manifest. 
The absence of a dissipative zone is apparent in Fig. \ref{FIG:Leith}.a, when compared both with the $\alpha$V-EDQNM case (see Fig. \ref{Fig:EDQNM}.a) and the Euler Voigt-$\alpha$ model (see Fig.\ref{fig:EulerSpectra2048}). This is certainly due to the locality in Fourier space of the $\alpha$V-Leith model.

As in the previous models, we look for self-similar solution of \eqref{Eq:LeithModelGen}. Introducing \eqref{Eq:SelfSimForm} in to \eqref{Eq:LeithModelGen} we obtain in the limit of $k\gg k_\alpha$ for the self-truncation exponent 
\begin{equation}
\eta=\frac{1}{ \beta r -1} \label{Eq:Eta_Leith}.
\end{equation}
This prediction coincides with the one of EDQNM for $r=2+\gamma/\beta$ (see Eq.\eqref{Eq:Eta_EDQNM}). The self-similarity behavior of $E_\alpha(k,t)$ is apparent in Fig.\ref{FIG:Leith}.b, where the self-similar form is displayed for $\beta=4$ and $r=2$. The inset shows the temporal evolution of $\kst(t)$ for different values of $\beta$. A power-law growth is manifest. The measured values of the exponent $\eta$ presented inTable \ref{tab:eta} are in good agreement with the prediction \eqref{Eq:Eta_Leith}.

The self-similar analysis leads to a non-linear second-order ordinary differential equation for $\Psi(z)$ (see \eqref{Eq:SelfSimForm}). This equation cannot be solved analytically but an asymptotic analysis predicts $\Psi(z)\sim z^2$  for $z\ll1$ and $\Psi(z)\sim ({\rm cte}-z)^{3/2}$ for $z\gtrsim1$. Data is compatible with this result (not shown). Note that, unlike the Euler-Voigt-$\alpha$ and $\alpha$V-EDQNM models, the Leith model presents a sharp cut-off instead of an exponential decay.

\section{Conclusion}\label{sec:conclusion}

In summary the Euler-Voigt-$\alpha$ model allowed us to show that its self-truncation regime reproduces the behavior of the truncated Euler equation \cite{Cichowlas:2005p1852}. We also found evidence for self-similarity in the long-time behavior of the energy spectrum. Introducing two different simplified models, the $\alpha$V-EDQNM model and the $\alpha$V-Leith model, we were able to show that they present behaviors similar to that of the Euler-Voigt-$\alpha$  model. We were able to determine the analytical values of the self-similar exponents of the simplified models.

In the present work we have used only integer values for $\beta$. As was noted in Sect.\ref{sec:EVdef}, choosing $\beta = 11/3$ yields an absolute equilibrium $E(k) \sim k^{-5/3}$ and, in two-dimensions, $\beta = 2/3$ also yields Kolmogorov scaling. This can represent an interesting alternative to the fractal decimation method that was used in reference \cite{PhysRevLett.108.074501}.

The present work can be naturally extended to the $2D$ and $3D$ Ideal MHD equations. In this context, it was recently shown that dynamo action can be triggered by turbulence in absolute equilibrium \cite{Ganga}. Thermalization with $\beta=11/3$ would allow for a more realistic velocity spectrum, mimicking the infinite Reynolds limit. 

As already stated in the introduction, sevral \emph{dissipative} Navier-Stokes-Voigt-$\alpha$ regularizations have been proposed as subgrid-scale models of classical turbulence closure problems both in hydrodynamics \cite{CaoTiti2009} and in magnetohydrodynamics \cite{Mininni2005}. We now end this paper with a few remarks on the relation of the present work with these more standard subjects.

First, all the calculations presented above were performed \emph{without dissipation}. This methodology clearly poses a challenge when one is interested in the Reynolds number ($Re$) of the flows, as {\it e.g.} in the implicit Large Eddie Simulations of reference \cite{Zhou2014}. However, in fully developped turbulence \cite{frischbook} the dissipative scale $k_d$ is related to the integral scale $k_I$ by $k_d\sim k_I Re^{3/4}$. Thus a first guess for a Reynolds number could be $Re\sim (k_d/k_I)^{4/3}$, with $k_d$ given by Eq.(\ref{eq:kd}) above (see section \ref{sec:st}).This important problem is beyond the present work and clearly deserves further attention.

A second question is how the sweeping and straining processes (for review articles, see {\it e.g.} \cite{Zhou2004,Zhou2010}) need to be modified in the Euler-Voigt-$\alpha$ and related fluid models. This problem was discussed, for the $\alpha$-Voigt EDQNM model, 
in the paragraph following Eq.\eqref{eq_eta}. The identification of sweeping and straining processes in the complete Euler-Voigt-$\alpha$ model Eq.(\ref{eq:Euler-VoigtG}) is an important problem that must also be left for further studies.

{\bf Acknowledgements}: 
We acknowledge useful scientific discussions with Claude Bardos, Uriel Frisch, Annick Pouquet, Samriddhi S. Ray  and Edriss Titi. The computations were carried out at IDRIS  and the M\'esocentre SIGAMM hosted at the Observatoire
de la C\^ote d'Azur.

\bibliographystyle{apsrev}

\bibliography{bibli_ReSub}

\end{document}